\documentclass{mem}
\usepackage{natbib}\usepackage{txfonts}\usepackage{balance}
\usepackage{graphicx}
\usepackage[a4paper,breaklinks,dvipdfm]{hyperref}
\idline{75}{282}
\begin{document}

\title{
Multiple populations in Galactic globular clusters: a survey in the Str\"omgren system
}

   \subtitle{}

\author{
J.\, Alonso-Garc\'ia\inst{1} 
\and M.\, Catelan\inst{1,2}
\and P.\, Amigo\inst{3}
\and C.\, Cort\'es\inst{4}
\and C. A.\, Kuehn\inst{5}
\and F.\, Grundahl\inst{6}
\and G.\, L\'opez\inst{7}
\and R.\, Salinas\inst{8}
\and H. A.\, Smith\inst{9}
\and P. B.\, Stetson\inst{10}
\and A. V.\, Sweigart\inst{11}
\and A. A. R.\, Valcarce\inst{12}
\and M.\, Zoccali\inst{1}
          }

  \offprints{J. Alonso-Garc\'ia}

\institute{
Departamento de Astronom\'{i}a y Astrof\'{i}sica, 
Pontificia Universidad Cat\'{o}lica de Chile, 
Av. Vicu\~na Mackenna 4860, 782-0436 Macul,
Santiago, Chile
\and
The Milky Way Millennium Nucleus,
Santiago, Chile
\and
Universidad de Valpara\'iso,
Valpara\'iso, Chile
\and
Universidad Metropolitana de Ciencias de la Educaci\'on,
Santiago, Chile
\and
Sydney Institute for Astronomy, University of Sydney, 
Sydney, Australia
\and 
Aarhus University,
Aarhus, Denmark
\and
KU Leuven,
Leuven, Belgium
\and
Finnish Centre for Astronomy with ESO, University of Turku,
Turku, Finland
\and
Michigan State University, 48824,
East Lansing, MI
\and
Dominion Astrophysical Observatory,
Victoria, Canada
\and
NASA Goddard Space Flight Center,
Greenbelt, MD
\and
Universidade Federal do Rio Grande do Norte,
Natal, Brazil
 \\
\email{jalonso@astro.puc.cl}
}

\authorrunning{Alonso-Garc\'ia}

\titlerunning{Multiple populations in Galactic GCs}

\abstract{ We are coming to believe that stellar populations in
  globular clusters are not as simple as they were once thought to
  be. A growing amount of photometric and spectroscopic evidence 
  shows that globular clusters host at least two
  different stellar populations. In our contribution to these
  proceedings we present the first results of a survey we are
  conducting to look for the presence of multiple populations in a
  significant number of Galactic globular clusters, using the
  Str\"omgren system. We intend to photometrically separate these
  populations and characterize their radial distributions and
  extensions.

   }
\maketitle{}

\section{Introduction}
Variations in the light element composition among the brightest stars
of some Galactic globular clusters (GCs) have been known for several
decades now (\citealt{fr81,no83,kr94}, and references therein). But
only recently, thanks to the results of extensive high-resolution
spectroscopic surveys in many
Galactic GCs \citep{ca09,jo12,co12,mu12}, have we found this to be
the rule in these objects. The presence of such star-to-star chemical
differences in GCs is not restricted to the brightest, most evolved
stars. Non-evolved, main-sequence and turn-off stars also show these
variations \citep{br94,gr01}, which suggests a primordial origin. The
most extended explanation suggests a self-enrichment scenario with at
least two star-formation episodes, where stars from later generations
are chemically enriched with respect to the first generation
(\citealt{va11}, and references therein), but the mechanism of
self-enrichment and its extension is still a matter of current
discussion and debate (\citealt{gr12}, and references therein).

Photometry has also proved to be very useful in tackling the problem
of disentangling the multiple populations of GCs, by observing and
examining their features in the color-magnitude diagram (CMD), such as
the presence of several red giant branches (RGBs; e.g., NGC 288 -
\citealt{ro11}; NGC 1851 - \citealt{ha09}), several horizontal
branches (HBs; e.g., Terzan 5 - \citealt{fe09}), several subgiant
branches (SGBs; e.g., NGC 1851 - \citealt{mi08}; NGC 362, NGC 5286,
NGC 6656, NGC 6715 and NGC 7089 - \citealt{pi12}), or even several
main sequences (MSs; e.g., NGC 2808 - \citealt{pi07}). Photometry
provides a means to increase enormously the limited spectroscopic
sample size, covering the whole population of the brightest stars in
the GC in much shorter times than spectroscopy, and reaching dimmer stars
in the more crowded, more central GC environments.  Observations using
the Str\"omgren filters, especially the ones in the ultraviolet
wavelength range, have been suggested to be more efficient in showing
the effects of different stellar populations, due to sensitivity of
their passbands to strong molecular bands such as CN, NH, or CH (e.g.,
\citealt{gr02,yo08,sb11,ca11}).

\section{Our survey}
We have recently started a survey of Galactic GCs, searching for the
photometric imprints of the separate populations present in them. To
carry out our observations, we are using the SOI camera installed in
the 4.1m SOAR telescope, located in Cerro Pach\'on, Chile. The SOI
field of view (FOV) is $5.25'\times5.25'$, with a pixel scale of
$0.154''$. This FOV is too small to cover the whole field of the
clusters, so for every cluster we are performing several different
pointings, from their centers to their outer regions. We are using
four Str\"omgren filters ($u$, $v$, $b$, and $y$), plus the Bessel $I$
for a more complete wavelength coverage. So far we have been able to
observe 30 Galactic GCs. We obtained the PSF photometry from the
images using an updated version of Dophot \citep{sc93,al12}. We are
calibrating the photometry using a set of GCs with previous
well-calibrated Str\"omgren photometry \citep{gr99}, and \citet{st00}
photometric standard stars in $I$. Also we have astrometrized our
observations by comparison with bright stars obtained in each field
from the Two Micron All Sky Survey (2MASS; \citealt{sk06}) catalog.

\begin{figure*}[t!]
\resizebox{\hsize}{!}{
\includegraphics[clip=true]{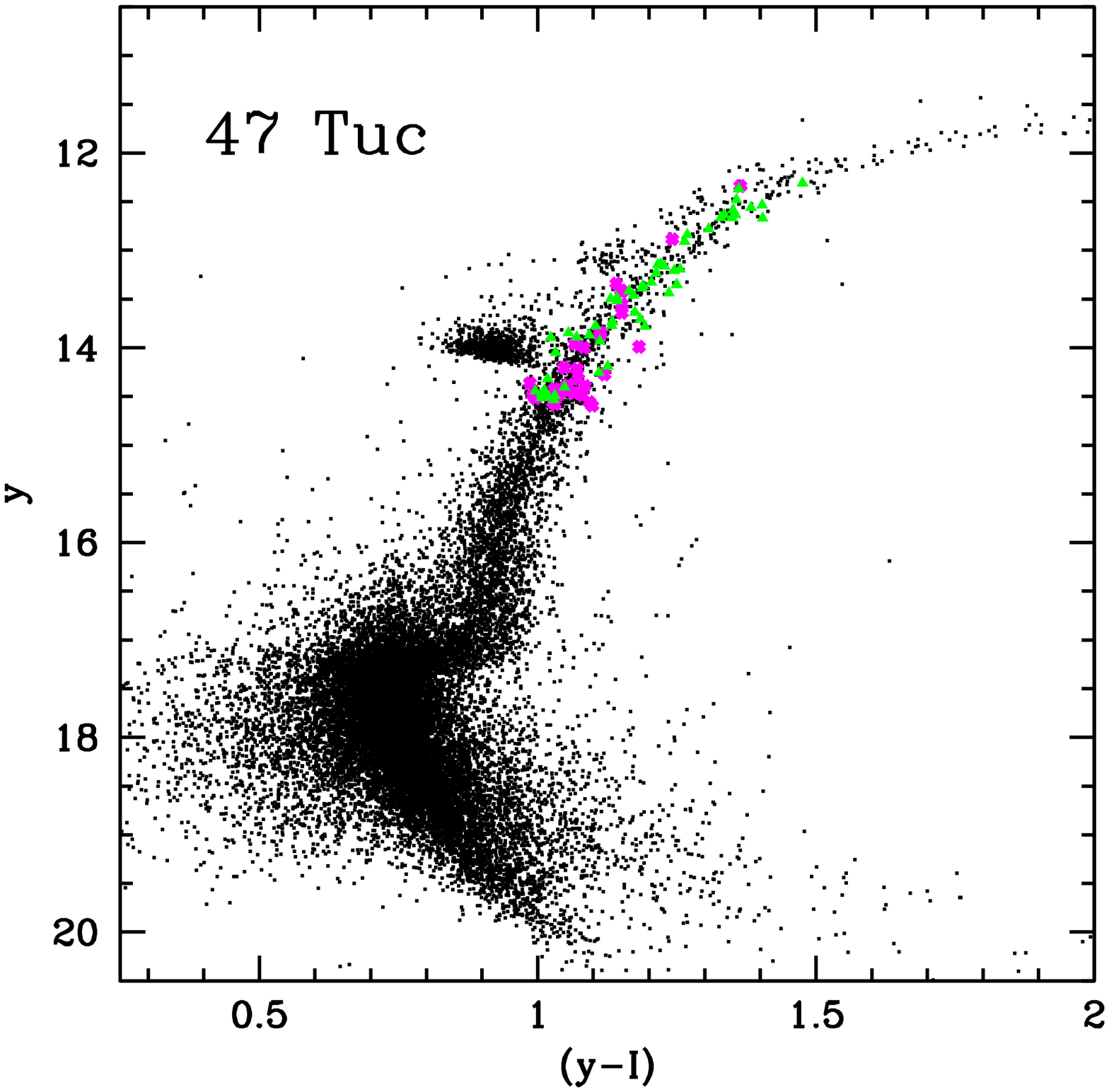}
\includegraphics[clip=true]{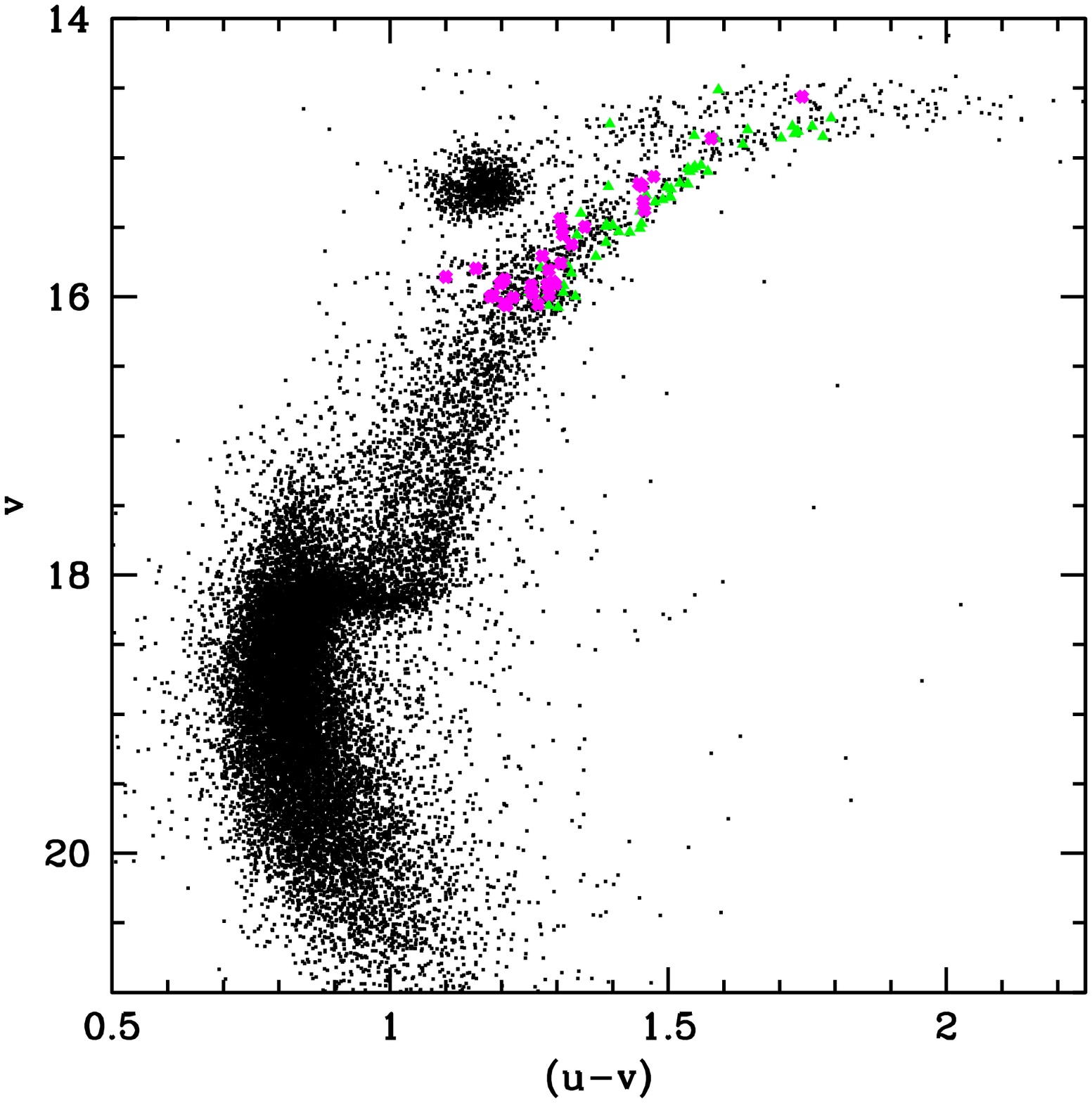}}
\resizebox{\hsize}{!}{
\includegraphics[clip=true]{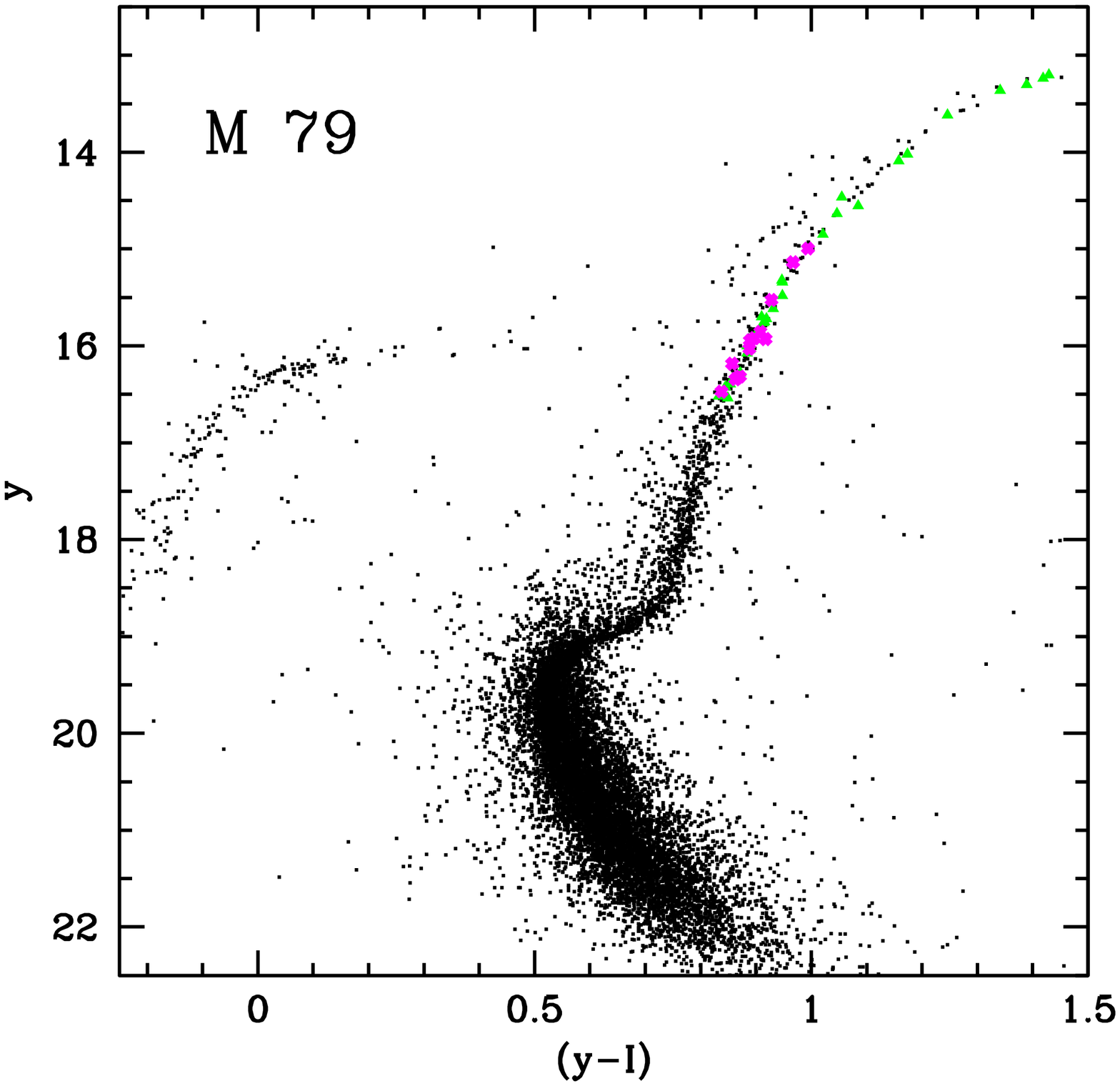}
\includegraphics[clip=true]{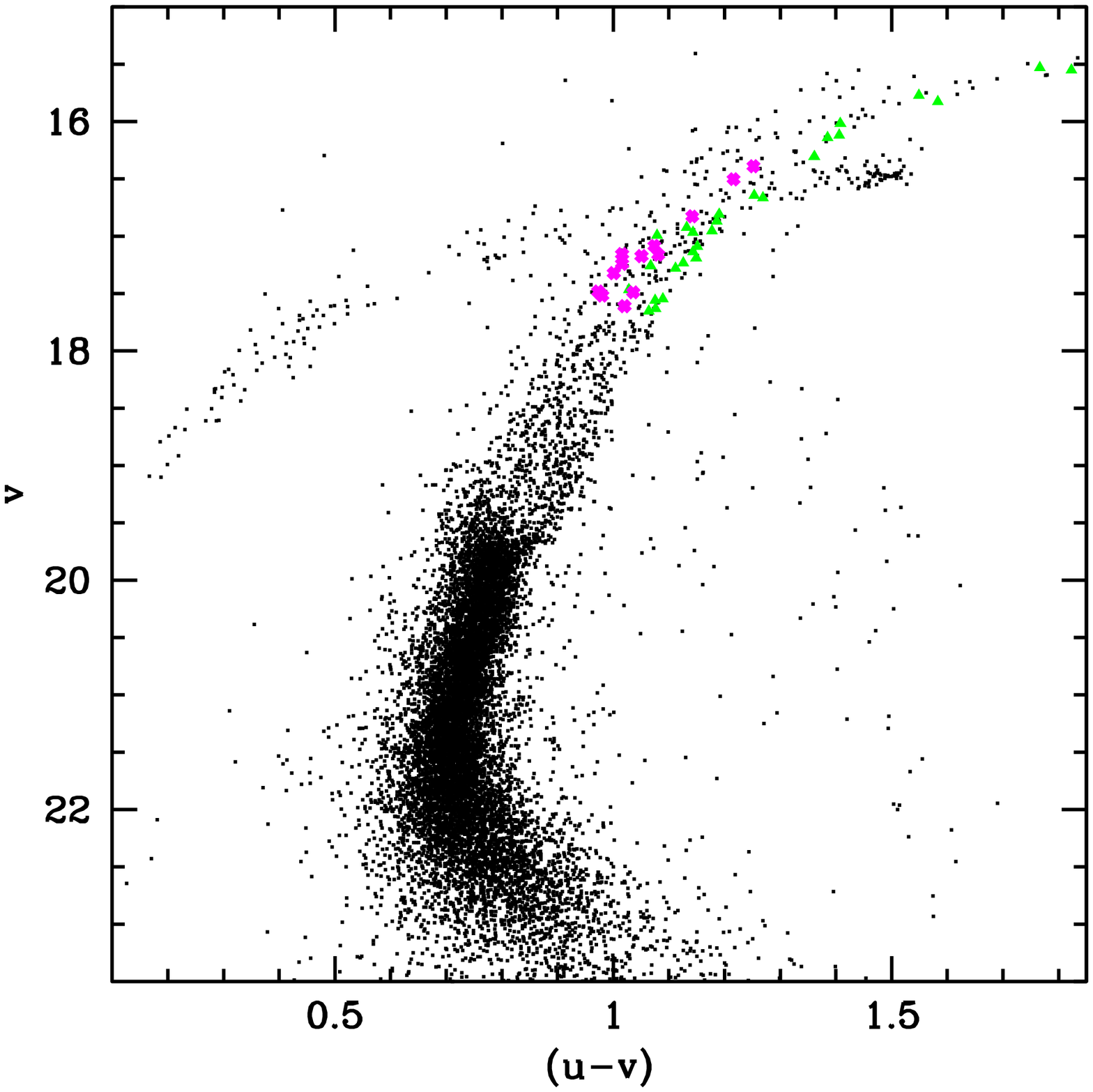}}
\caption{
\footnotesize
CMDs for 47~Tuc (upper panels) and M~79 (lower panels). Despite the difference in metallicities between both clusters ([Fe/H]=-0.72 vs. [Fe/H]=-1.60; \citealt{ha96}), we observe for both broadenings on the RGB sequences in our bluest filters (right panels) that are not present in our reddest filters (left panels). Magenta crosses and green triangles represent stars from primary and secondary populations as defined spectroscopically by \citet{ca10}. While in our reddest filters these different populations are mixed, they are clearly correlated with the photometric separation observed using our bluest filters.
}
\label{figrgb}
\end{figure*}

\begin{figure*}[t!]
\resizebox{\hsize}{!}{
\includegraphics[clip=true]{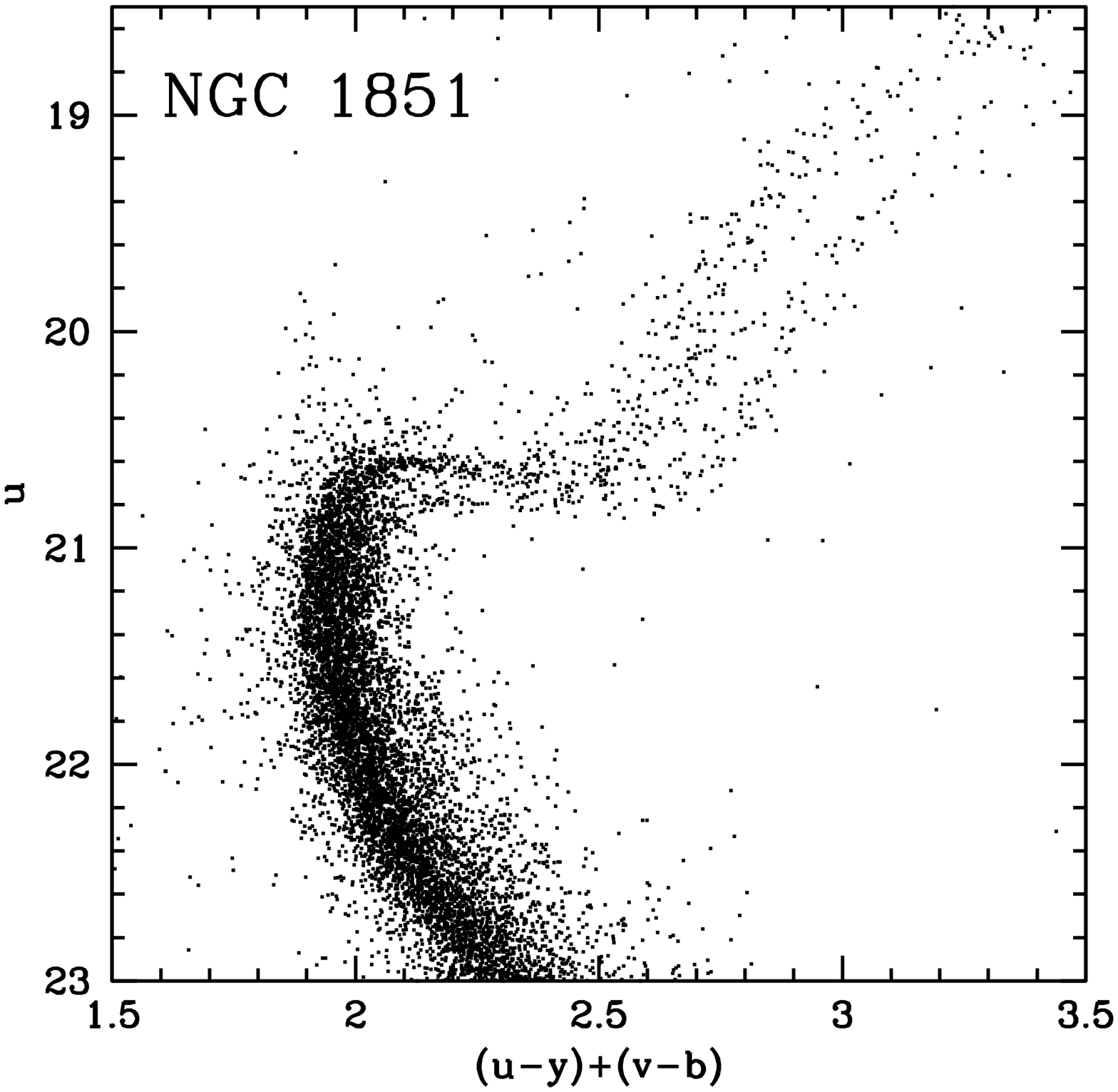}
\includegraphics[clip=true]{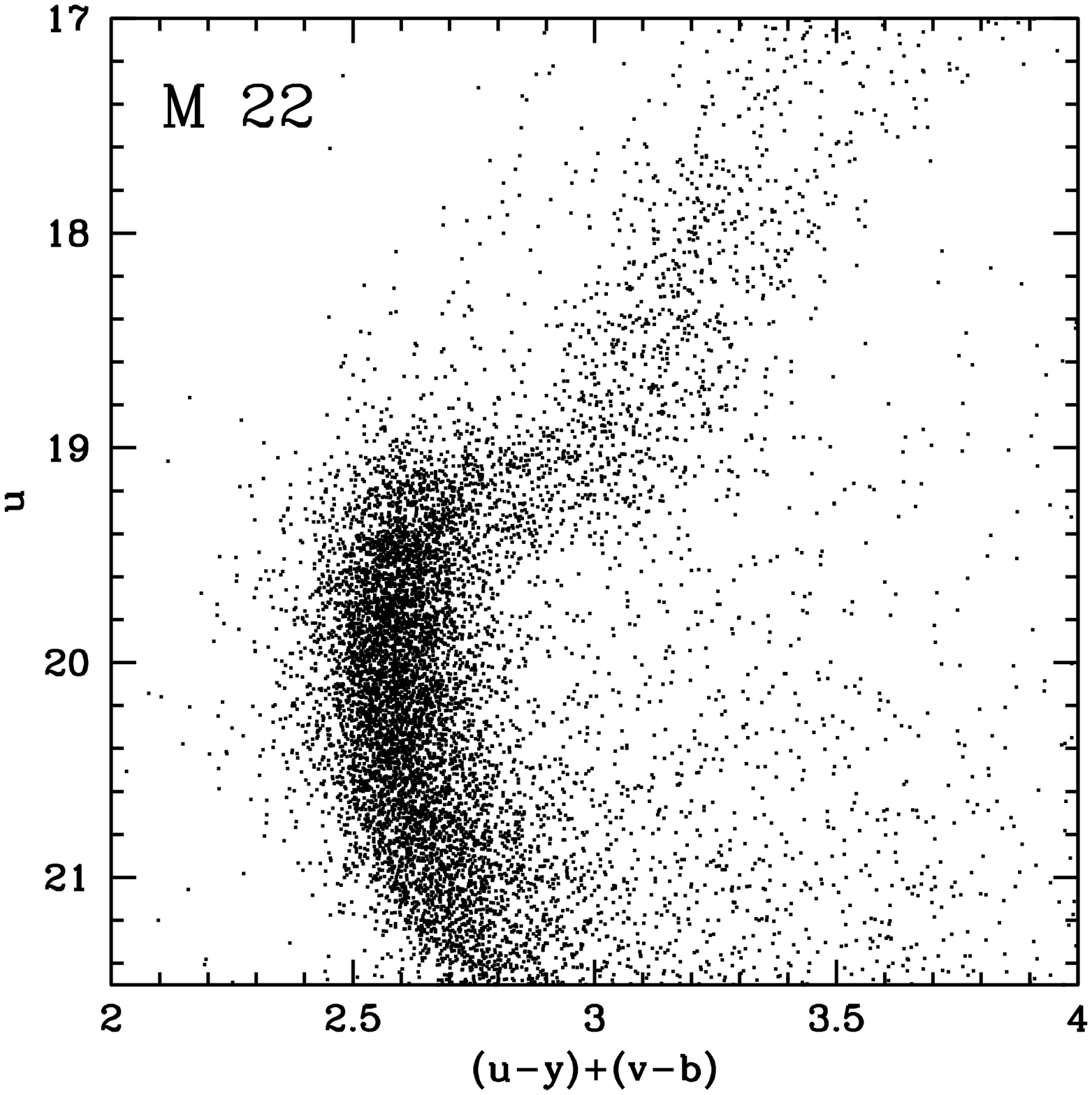}}
\caption{
\footnotesize
CMDs for NGC~1851 (left panel) and M~22 (right panel). The separation in the populations is clearly visible  in the SGB and RGB using the color index $(u-y)+(v-b)$ introduced by \citet{la12}.
}
\label{figsgb}
\end{figure*}

We have generated CMDs for the GCs in our sample for the different
available magnitudes and color-indices. We have found that for our
reddest color-indices we obtain CMDs with narrow, well-defined
evolutionary sequences (see figure \ref{figrgb}, left panels). These
CMDs, while adequate to obtain information about the global GC parameters
(e.g., distance, age, reddening), do not allow us to separate stars into
different populations. On the other hand, the CMDs generated with our
bluest color-indices, especially those that contain the ultraviolet
$u$ passband, show significant broadenings in their RGBs (see figure
\ref{figrgb}, right panels). When we overplot on our bluest CMDs the
stars that have been spectroscopically separated in different
populations \citep{ca10}, we observe that stars from different
populations group in different sides of the RGB, with only a few
exceptions (e.g., NGC~2808, NGC~7078). This is a clear indication of a
correlation among spectroscopic and photometric separations. A few of
the clusters in our sample also show broadenings and separations
in their SGBs (see figure \ref{figsgb}; \citealt{mi08,pi12}). These
photometric separations have been associated with the small percentage
of clusters that show variations in Fe and s-process elements
\citep{ma09,mi12}.

\section{Summary}
We are currently conducting a photometric survey of a significant
sample of Galactic GCs in the Str\"omgren system using the 4.1m SOAR
telescope. We are searching for signatures of multiple stellar
populations. First results are encouraging, showing a clear
correlation between the broadenings we find in the evolutionary
sequences of the sampled clusters' CMDs and the spectroscopic
separations reported in the literature. We aim to disentangle these
populations, and to study their radial distribution and proportion
ratios from the cluster center out to the outskirts of every cluster
in our sample.

\begin{acknowledgements}
  This project is supported by the Chilean Ministry for the Economy,
  Development, and Tourism's Programa Iniciativa Cient\'ifica Milenio
  through grant P07-021-F, awarded to The Milky Way Millennium
  Nucleus; by Proyecto Fondecyt Postdoctoral 3130552; by Proyecto
  Fondecyt Regular 1110326; by Proyecto Basal CATA PFB-06; and by Anillos
  ACT-86.
\end{acknowledgements}

\bibliographystyle{aa}

\end{document}